\documentclass[twocolumn,preprintnumbers,aps,amsmath,amssymb]{revtex4}
\usepackage{latexsym}
\usepackage{graphics}

\begin{document}

\title{A new type  fit for the  CKM  matrix elements }
\author{Petre Di\c t\u a  } \email{ dita@zeus.theory.nipne.ro}

\affiliation{ Institute of Physics and Nuclear Engineering,
P.O. Box MG6, Bucharest, Romania}
\date{\today}

\begin{abstract}
The aim of the paper is to propose a new type of fits in terms
 of invariant quantities for finding the entries of the CKM matrix from 
the quark sector, by using the mathematical solution  to the reconstruction
 problem of $3\times 3$ unitary matrices from experimental data, recently found.  The necessity of this type of fit comes from the compatibility conditions 
between the data and the theoretical model formalised by the CKM matrix, 
which  imply  many strong nonlinear conditions on moduli which all have
 to be satisfied in order to obtain a unitary matrix.\end{abstract}


\maketitle

\section{Introduction}
In the description of the electroweak interactions within the Standard Model a fundamental r\^ole  is played by the unitary   CKM matrix, parametrised by  four independent parameters:  three mixing angles, $\theta_{ij}$, and a  $CP$ violating phase, $\delta$. On the other hand from experiments one finds numerical values for other quantities such as the  moduli, or the angles of so called unitarity triangles, and by consequence  the reconstruction of  unitary   matrices from experimental data is a central problem in the electroweak interactions. Only recently a  mathematical solution  to this problem was found, \cite{Di}, and one aim of the paper is to explore its implications upon the electroweak phenomenology. One  conclusion of the above cited paper was that it is necessary to provide phenomenological models for analysing  data in terms of rephasing invariant quantities, i.e. the CKM matrix moduli, the Jarlskog invariant, $J$, \cite{JS}, the area of any unitarity triangle, \cite{BL}, or the  phases of \,its  complex entries \cite{AKL}. In this paper we will choose the moduli as independent parameters because with this choice the reconstruction of a unitary matrix from data is essentially unique, see Ref.\,\cite{Di}. In the same time this choice allows us to define  phenomenological models which lead to  sets of non linear constraints which have to be satisfied by data in order to come from, or to be compatible with a unitary matrix. Hence the data could be compatible with unitarity, or even could disprove the  CKM unitary model, i.e.  nobody guarantees us  that  the moduli $|U_{ij}|$, the mixing  angles $\theta_{ij}$, or the phase $\delta$  extracted from  experiments come from a unitary matrix.
In phenomenological analyses it is usually  assumed that irrespective how the measured data are, they are compatible to the existence of a unitary matrix, and in  literature  one finds  statements such as: 
the most stringent test of CKM unitarity is the relation
 \begin{eqnarray}
V_{ud}^2+V_{us}^2+V_{ub}^2=1\label{cm}\end{eqnarray}
see, e.g., Ref.\,\cite{BLI}, where $V_{ij}$ are the measured moduli values, statement which is  not true, because there is a  natural embedding of unitary matrices  into a larger class of matrices, that of double stochastic ones, \cite{MO}, and we have to find methods to discriminate between the two sets.

A $3\times 3$  matrix $M$ is said to be  double stochastic if its elements satisfy the relations 
\begin{eqnarray} m_{ij}\ge 0,\quad \quad\sum_{i=1}^3\,m_{ij}=1,\quad\quad \sum_{j=1}^3 m_{ij}=1\label{ds}\end{eqnarray}

 The unistochastic matrices, \cite{MO}, are a subset of the double stochastic ones defined by
\begin{eqnarray} m_{ij}= |U_{ij}|^2\label{ds1}\end{eqnarray}
where $U$ is a $3\times 3$  unitary matrix. Hence even if all the six relations such as (\ref{cm}) are exactly satisfied the corresponding matrix could be non-unitary.

The double stochastic matrices have an important property, they are a convex set, i.e., if $M_1$ and $M_2$ are double stochastic, so  is their convex combination $\alpha\,M_1+(1-\alpha)M_2$, $\alpha \in [0,1]$, as it is easily checked. This property is very important because it was the necessary ingredient  to devise a method for doing statistics on unitary matrices.

The relations (\ref{ds}) together with the embedding relations (\ref{ds1}) define a phenomenological model which allows to find formulae for the mixing angles $s_{ij}=\sin\theta_{ij}$ and  $\cos\delta$ in terms of four independent moduli $V_{ij}$, and the compatibility condition between data and unitarity property leads to the most constraining unitarity condition, namely, $\cos\delta$ as function of  moduli should take physical values, i.e. $-1\le \cos\delta\le 1,$ see Ref.\,\cite{Di}. 

The choice of four moduli as independent parameters is very appealing from a theoretical point of view, since, as we said before, in this case the reconstructed unitary matrix is essentially unique. On the other hand it is naturally to assume that the physical results of any phenomenological analysis of experimental data must be invariant with respect to the choice of the unitary matrix form, e.g., Kobayashi-Maskawa, \cite{KM}, or Chau-Keung, \cite{CK}, form. Although both the above cited parameterisations depend on mixing angles, $\theta_{ij}$ and $CP$ non-conserving phase $\delta$, only the last parameter is invariant, the numerical values for $\theta_{ij}$ depend on the chosen form. The mixing angles being not invariant quantities, it is better to avoid their use in phenomenological analyses. If the physical results are not invariant, this could be a signal that the unitary model is not compatible with experimental data, or that our phenomenological data analysis  is done with wrong parameters.

 The above considerations  imply  that for defining a {\em novel type of CKM fits} we have to make clear a few natural assumptions, which we will state as  axioms. Concerning the experimental side  an obvious  axiom could be the following:
\\ {1)\,\,\em the numerical values for all the measured moduli,  $|U_{ij}|,$ should be the same irrespective of the physical processes used to determine them}.

The second axiom comes from the constraints imposed by data on the explicit theoretical tool used for doing phenomenological analyses. Thus the second axiom could be:\\{2) \,\,\em the physical results obtained from data analyses must be invariant with respect to the choice of four independent invariant quantities used to parametrise the data}. 

Because in this paper  we will  use moduli as invariant parameters, and the number   of four independent moduli groups equals 58, one gets 165 different forms for $\cos\delta$, which when computed by using real data one finds different values, instead of a single one. Thus the above  axiom will have in this paper the form:\\
{2a) \,\,\em the physical results must be invariant with respect to all the choices of four independent moduli groups used to parametrise the data}. 

  Of course for defining phenomenological models and for a full  reconstruction of  unitary matrices from error affected data we have to use one explicit form of the CKM matrix, and in the following we use the form from Ref.\,\cite{CK}, which, by making full use of its invariance at multiplication at left and at right with diagonal phases matrices, we write it as 
\begin{widetext}
\begin{eqnarray}
U=\left(\begin{array}{ccc}
c_{12}c_{13}&c_{13}s_{12}&s_{13}\\
-c_{23}s_{12}e^{i \delta}-c_{12}s_{23}s_{13}&c_{12}c_{23}e^{i \delta}-s_{12}s_{23}s_{13}&s_{23}c_{13}\\
s_{12}s_{23}e^{i\delta}-c_{12}c_{23}s_{13}&-c_{12}s_{23}e^{i \delta}-s_{12}c_{23}s_{13}&c_{23}c_{13}
\end{array}\right)\label{ckm} 
\end{eqnarray}\end{widetext}
with the standard notation: $c_{ij}=\cos \theta_{ij}$ and  $s_{ij}=\sin \theta_{ij}$; $\theta_{ij}$,  $ij=12,\,13,\,23$, denote the  mixing angles and $\delta$ is the phase that encodes the  $CP$-violation in the electroweak sector.

The paper content is as follows. In Section 2 we define two phenomenological models that will provide the 
 necessary and sufficient conditions the data have to satisfy in order that they  should come from a unitary matrix.  Eventually by finding that data come from a unistochastic matrix we have to provide  an algorithm for the reconstruction of  $U$ from the error affected data. In Section 3 we test the moduli data available from experiments, lattice computations, and global fits. We show that the experimental errors are quite large so there exists a continuum of unitary matrices compatible with data.
 The paper ends by Conclusion.


\section{Phenomenological models}
By phenomenological model we understand  a relationship between the entries of a unitary matrix and the measured quantities, such as the  moduli.
Hence in the following  we assume the knowledge from experiment of the  moduli of unitary matrix entries such as   (\ref{ckm}), which we write as the entries of a positive  matrix 
\begin{eqnarray}
V=\left(\begin{array}{ccc}
\vspace*{1mm}
V_{ud}^2&V_{us}^2&V_{ub}^2\\
\vspace*{1mm}
V_{cd}^2&V_{cs}^2&V_{cb}^2\\

V_{td}^2&V_{ts}^2&V_{tb}^2\\
\end{array}\right)\label{pos}
\end{eqnarray}
 For the current  state of the art concerning the determination of the above quantities see, e.g., Refs.\cite{BLI} and \cite{WM}. As the notation suggests  we make a clear distinction between the elements of the unitary CKM matrix $U$ and the {\em positive} entries  matrix $V$ provided by  data. In other words we make a distinction between the theoretical quantities $U_{ij}$ and the experimental moduli $V_{ij}$, although in an ideal situation the relation $|U_{ij}|=V_{ij}$ will hold.

 The main  phenomenological  problem is to see in what conditions  from a matrix such as (\ref{pos}) one can reconstruct a unitary matrix as (\ref{ckm}), i.e. to see if the  data are compatible with the theoretical model. The compatibility conditions are better understood in  the frame of a phenomenological model, which in our case will be  a relationship between the theoretical object (\ref{ckm}) and the experimental data (\ref{pos}).  The theoretical tool we have at our disposal for defining such a  relationship is the unitarity property of the CKM matrices. The aim of any  phenomenological analysis is at least twofold: a) checking the consistency of data   (\ref{pos}) with the theoretical model  (\ref{ckm}), and b) determination of theoretical parameters $s_{ij}$ and $\delta$ from the experimental data (\ref{pos}) if the data and the theoretical model are compatible. 
In other words our scope is the reconstruction of the unitary matrix (\ref{ckm}) from experimental data. In the following we present two phenomenological models: unitarity condition method and unitarity triangle method.

In both the models we use the embedding of unitary matrices into the set of double stochastic matrices, see (\ref{ds1}), implying that  the following  relations should hold
\begin{eqnarray}
\sum_{i=d,s,b} |U_{ji}|^2-1=\sum_{i=d,s,b} V_{ji}^2-1=0, \quad j=u,c,t\nonumber \\
\sum_{i=u,c,t} |U_{ij}|^2-1=\sum_{i=u,c,t} V_{ij}^2-1=0, \quad j=d,s,b \label{sto}
\end{eqnarray}
In the above relations $V_{ij}$ are numbers affected by errors obtained from experiments, and the exact fulfilment of Eqs.(\ref{sto}) is not at all guaranteed in any phenomenological analysis. Even if they are exactly satisfied, the unitarity property  could  not be satisfied. And this is one main reason for using different notations for  experimental moduli $V_{ij}$, and for theoretical moduli $|U_{ij}|$. The distinction is also necessary since the phenomenologists are not  aware of the natural embedding of unitary matrices, through  relation (\ref{ds1}), into a larger set, that of  double stochastic matrices (\ref{ds}), i.e. they are not yet aware of the necessity to find a criterion for the separation  of unitary matrices within the double stochastic set, only the intersection of the two sets being relevant for the electroweak physics phenomenology.

\subsection{Unitarity condition method}
The specific condition defining this model is the fulfilment of the relation
\[V=|U|^2\]
   between the theoretical object (\ref{ckm}) and the experimental data (\ref{pos}),
relation that has to be understood as working entry wise  leading to the following relations
\begin{eqnarray}
V_{ud}^2&=&c^2_{12} c^2_{13},\,\, V_{us}^2=s^2_{12}c^2_{13},\,\,V_{ub}^2=s^2_{13}\nonumber \\
 V_{cb}^2&=&s^2_{23} c^2_{13},\,\,
 V_{tb}^2=c^2_{13} c^2_{23},\nonumber\\
V_{cd}^2&=&s^2_{12} c^2_{23}+s^2_{13} s^2_{23} c^2_{12}+2 s_{12}s_{13}s_{23}c_{12}c_{23}\cos\delta,\nonumber\\
V_{cs}^2&=&c^2_{12} c^2_{23}+s^2_{12} s^2_{13} s^2_{23}-2 s_{12}s_{13}s_{23}c_{12}c_{23}\cos\delta,~~~~~\label{unitary}\\
V_{td}^2&=&s^2_{13}c^2_{12}c^2_{23}+s^2_{12}s^2_{23}-2 s_{12}s_{13}s_{23}c_{12}c_{23}\cos\delta\nonumber,\\
V_{ts}^2&=&s^2_{12} s^2_{13} c^2_{23}+c^2_{12}s^2_{23} +2 s_{12}s_{13}s_{23}c_{12}c_{23}\cos\delta\nonumber
\end{eqnarray}
 
 This phenomenological model is similar to that proposed by Wolfenstein, \cite{W}, i.e. it is a direct relationship between the measured values $V_{ij}$ and the theoretical parameters entering the unitary matrix (\ref{ckm}), $s_{ij}$ and $\delta$. The difference between the two approaches is  the following: we make no approximations on the right hand of Eqs.\,(\ref{unitary}), and we  will explicitly make  use of the double stochasticity relations (\ref{sto}).  One sees also  that on the left hand in Eqs.\,(\ref{unitary}) there are sets of {\em nine} numbers, $V_{ij}^2$, obtained from experiments, and on the right hand enter only {\em four} independent parameters. Hence  the consistency problem  of the  equations (\ref{unitary}) is a natural one, and it has to be resolved.
 An other remark is that the  $\delta$ dependence of equations (\ref{unitary}) is through the cosine function, which is an even function, so we can restrict the range of $\delta$ to $(0,\pi)$, when we use as independent parameters the moduli $V_{ij}$, without loss of generality.

The main  problem to be solved is to find in what conditions the system (\ref{unitary}) could have a physical solution, for   arbitrary numbers $V_{ij}$ satisfying Eqs.(\ref{sto}), i.e. from the full set of double stochastic matrices. In Ref. \cite{Di} it was shown that the necessary and sufficient conditions 
the data have to satisfy in order to  the matrix (\ref{pos}) comes from a unitary matrix are 
\begin{eqnarray} 0\le s_{ij}\le1, \quad {\rm and}\quad -1\leq\cos\delta\leq 1\label{unit1}\end{eqnarray} 
when $s_{ij}$ and $\cos\delta$ are found by resolving the equation system (\ref{unitary}).
 The  most constraining  condition is the second one, i.e. $\cos\delta\in (-1,1)$, which is the separation criterion between the double stochastic and unitary matrices, whose fulfilment is compulsory. In fact for the $3\times 3$ matrices, if the relations (\ref{sto}) are satisfied, there always exists a solution for $s_{ij}$ which is physical, and, e.g., it could be  obtained by using three independent relations from the first five appearing in equations (\ref{unitary}). The relations (\ref{unit1}) are the consistency conditions between the data and the theoretical model.

To see the constraining power of the above relations (\ref{unit1}) we assume for the moment that the relations (\ref{sto}) are exactly satisfied. The matrix (\ref{ckm}) depends on {\em four} independent parameters and we choose them as four independent moduli, i.e. four $|U_{ij}|$. Now we use the relations (\ref{sto}) and the entries from the data set  (\ref{pos}) to construct a matrix, $S=(S_{ij})$, whose square entries form a double stochastic matrix. In order to simplify the formulae form we make the notation
\begin{eqnarray}
V_{ud}&=&a,\quad V_{us}=b,\quad V_{ub}=c\nonumber\\
V_{cd}&=&d,\quad V_{cs}=e,\quad V_{cb}=f
\end{eqnarray}
For example, if the four independent parameters are the moduli $V_{us} = b,\,V_{ub} = c,\,\, V_{cd} = d,\,\, V_{cb} = f$, the matrix $S^2=(S_{ij}^2)$, where
\begin{eqnarray}
S&=&(S_{ij})=\label{toy1}\\
&&\left(\begin{array}{ccc}
\sqrt{1-b^2-c^2}&b&c\\
d&\sqrt{1-d^2-f^2}&f\\
\sqrt{b^2+c^2-d^2}&\sqrt{d^2+f^2-b^2}&\sqrt{1-c^2-f^2}
\end{array}\right)\nonumber
\end{eqnarray}
is double stochastic. The above choice is motivated by the fact that the  moduli $b,\,c$ and $f$ are the essential parameters used by the CKM fitter groups. By using the relations   $V_{ij}^2=S_{ij}^2$ in equations (\ref{unitary}) one gets values for
 $s_{ij}$ and $\cos\delta$ as follows
\begin{eqnarray}
&&s_{12}^{(1)}=\frac{V_{us}}{\sqrt{1-V_{ub}^2}}=\frac{b}{\sqrt{1-c^2}},\nonumber\\
&&  s_{23}^{(1)}=\frac{V_{cb}}{\sqrt{1-V_{ub}^2}}=\frac{f}{\sqrt{1-c^2}},\quad
 s_{13}^{(1)}=V_{ub}=c \label{coef}\\
&&\cos\delta^{(1)}=\nonumber\\*[2mm]
&& \frac{d^2(1-c^2)^2-b^2(1-c^2)+f^2(b^2-c^2+c^2(b^2+c^2))}{2 b c f \sqrt{1-b^2-c^2}\sqrt{1-c^2-f^2}}\nonumber
\end{eqnarray}
By doing  similar computations, but now with the independent parameters $a,\,\,b,\,\,d,\,\,e$,  entering in a double stochastic matrix similar to (\ref{toy1}), one gets  
\begin{widetext}\begin{eqnarray}
s_{12}^{(2)}=\frac{b}{\sqrt{a^2+b^2}},\quad s_{13}^{(1)}=\sqrt{1- a^2-b^2},\quad s_{23}^{(2)}=\frac{\sqrt{1-d^2-e^2}}{\sqrt{a^2+b^2}}~~~~~~~~~~~~~~~~~~~\nonumber\\
\cos\delta_2^{(2)}=\frac{b^2(1-b^2)-a^2(1-a^2)+ d^2(a^2-b^2+b^2(a^2+b^2))-e^2(a^2(a^2-1)+b^2(1+a^2))}{2\,a\,b\,\sqrt{1-a^2-b^2}\,\sqrt{1-d^2-e^2}\,\sqrt{a^2+b^2+d^2+e^2-1}\label{cod}}\end{eqnarray}\end{widetext}
and so on.
Now we provide two numerical matrices, built as\,  in (\ref{toy1}), whose squared moduli are double stochastic.
\begin{eqnarray}
M_1&=&\left(\begin{array}{ccc}
\frac{993907}{10^{6}}&\frac{22691}{10^{5}}&\frac{ \sqrt{17007251}}{10^{6}}\\*[2mm]
\frac{2269}{10^{4}}&\frac{24327}{25000}&\frac{3\sqrt{453251}}{50000}\\*[2mm]
\frac{\sqrt{21545351}}{10^{6}}&\ \frac{\sqrt{16271665}}{10^{5}}&
\frac{\sqrt{998351289149}}{10^{6}}\end{array}\right)\\*[2mm]
&\approx&\left(\begin{array}{ccc}
0.973907&0.22691&0.004124\\
0.2269&0.97308&0.040394\\
0.004642&0.040338&0.999175\end{array}\nonumber\right)
\end{eqnarray}

\begin{eqnarray}
M_2&=&\left(\begin{array}{ccc}
\frac{973907}{10^{6}}&\frac{22691}{10^{5}}&\frac{ \sqrt{17007251}}{10^{6}}\\*[2mm]
\frac{2267}{10^{4}}&\frac{9 \sqrt{7306909}}{25000}&\frac{3\sqrt{453251}}{50000}\\*[2mm]
\frac{\sqrt{112265351}}{10^{6}}&\frac{\sqrt{15364455}}{10^{5}}&\frac{\sqrt{998260569149}}{ 10^6 }\end{array}\right)\\*[2mm]
&\approx&\left(\begin{array}{ccc}
0.973907&0.22691&0.004124\\
0.2267&0.973127&0.04039\\
0.010596&0.039197&0.999175\end{array}\nonumber\right)
\end{eqnarray}

The moduli entering the matrices $M_1$ and $M_2$ have been chosen such that they should 
not be  too far from the recommended moduli values given by the PDG 2006 data, \cite{WM}, i.e. within the error corridors, and not too far from the central values of the last fit by CKM fitter Group \cite{JC}. Their difference is
\begin{eqnarray}
M_1-M_2\approx\left(\begin{array}{ccc}
0&0&0\\
0.0002&-4.7\times 10^{-5}&0\\
-0.00595& 0.00114&0\\
\end{array}\right)\label{dif}\end{eqnarray}
i.e. the entry moduli of the matrix (\ref{dif}) are much small than the current experimental errors on the corresponding moduli. In fact only the modulus $M_1(2,1)$ was changed by an amount of  $2\times 10^{-4}$ and this change was propagated to get again a double stochastic matrix.

By using  the formulae  (\ref{coef}) and (\ref{cod}) one gets  from $M_1$
\begin{widetext}
\begin{eqnarray}&&s_{12}^{(i)}=\frac{226910}{\sqrt{999982992749}}\approx0.226912,\qquad
s_{13}^{(i)}=\frac{\sqrt{17007251}}{10^6} \approx 0.004124,\nonumber\\
&&s_{23}^{(i)}=60\sqrt{\frac{453251}{999982992749}}\approx0.040395,\qquad\qquad\qquad i=1,2,...\label{u1}\\
&&\cos\delta^{(i)}=
\frac{613855254083724801174896537}{64224997110656250\sqrt{8008162692156206007509}}\approx 1.068\nonumber\end{eqnarray}\end{widetext}

The first remark is that $s_{ij}$ and $\cos\delta$ take the same values when the data come from an exact double stochastic matrix, i.e. they are independent of the four independent moduli we choose to parameterise the data. This is a consequence of the fact that the double stochasticity (unitarity) properties do not change if we interchange the columns and/or the rows between themselves, or we use the transposed, or the complex conjugated matrix.
The second remark is that, although the entries of the matrix $M_1$ are good from an experimental point of view, they are not not good from a theoretical point of view since $\cos\delta$ is unphysical.

From the second matrix, $M_2$, we get 
\begin{widetext}
\begin{eqnarray}s_{12}^{(i)}=\frac{226910}{\sqrt{999982992749}}\,,\quad s_{13}^{(i)}=\frac{\sqrt{17007251}}{10^6},\quad
s_{23}^{(i)}=60\sqrt{\frac{453251}{999982992749}}\,,\quad i=1,2,\dots\label{u2}\\*[2mm]
\cos\delta^{(i)}=-
\frac{1897412766859734762379196101}{128449994221312500\sqrt{8008162692156206007509}}
\approx-0.165~~~~~~~~~~~~~~~~\nonumber\end{eqnarray}\end{widetext}

From a theoretical point of view the second matrix, $M_2$, is compatible with unitarity. Please remark that in both the cases the mixing angles $s_{ij}$ take identical values although only $M_2$ comes from a unitary matrix. Hence the separation criterion between unitary matrices and purely double stochastic ones is provided by the physical values taken by $\cos\delta$. By looking at the difference matrix $M_1-M_2$, Eq.(\ref{dif}), one sees that the moduli differ by numbers which are smaller than the experimental errors and, in spite of this,  $\cos\delta$ has a substantial  jump from a unphysical value, 1.068, to a physical one, - 0.165, which is equivalent to $\delta=99.5^{\circ}$.
Thus the compatibility condition between  data and  theoretical model, 
$-1\le \cos\delta\le 1$, puts very strong conditions on  data in order to they  come from a unitary matrix, and, as the above example shows, the condition is very sensitive even to small moduli variation.  

When  the relations (\ref{sto}) are exactly satisfied the reconstruction algorithm of a unitary matrix from a double stochastic one  is the following: start with a double stochastic matrix as $M_1^2$, or $M_2^2$,  solve the equations (\ref{unitary}) and obtain  results as those given by equations (\ref{u1}) and (\ref{u2}). If the numerical value for $\cos\delta$ satisfies the inequalities (\ref{unit1}), then with the values for $s_{ij}$ and $\cos\delta$  go to equation (\ref{ckm}) and find the corresponding unitary matrix. In the above cases only the numerical results from $M_2$ are compatible to the existence of a unitary matrix, while those from the $M_1$ matrix are not, although the matrix  $M_1$ is a double stochastic one, and from the usual phenomenological point of view both  data could be considered as being  ``physical''. The reconstruction of the unitary matrix from the $M_2$ data is the matrix $\mathcal{U}$ which is obtained by the substitution of numerical values for $s_{ij}$ and $\cos\delta$  given by Eqs.\,(\ref{u2}) into formula (\ref{ckm}).

The true real case is when the double stochasticity relations (\ref{sto}) are not exactly satisfied, and by consequence the double stochastic matrices, built such as the matrix $S$, are different. 
 To see what happens in this case   we will use only  the central values from the  PDG 2006 data \cite{WM}, which are
\begin{eqnarray}
 a&=&0.97377,\quad b=0.2257,\quad  c=4.31\times 10^{-3},\label{m2}\\
 d&=&0.230,\quad ~~~e=0.957,\quad ~f=41.6\times 10^{-3} \nonumber\end{eqnarray}
and those from the fit \cite{JC1}, namely
\begin{eqnarray}
 a&=&0.97504,\quad b=0.2221,\quad  ~c=3.505\times 10^{-3},\label{m3}\\
 d&=&0.222,\quad ~~~e=0.97422,\quad f=40.8\times 10^{-3} \nonumber\end{eqnarray}
With them we form  the double stochastic matrices corresponding to the phenomenological models (\ref{coef}) and, respectively, (\ref{cod}), and one   gets in the first case
\begin{eqnarray} 
\cos\delta^{(1)}=&25.98,\quad\cos\delta^{(2)}=1.58\label{n1}
\end{eqnarray}
and 
\begin{eqnarray} 
\cos\delta^{(1)}=&0.6,\quad\cos\delta^{(2)}=-0.377\,i\label{n4}
\end{eqnarray}
in the second case. If in the first case the results could be considered as being ``normal'' since the moduli were obtained  by doing some weighted means on moduli data obtained by experimenters, in the second case the numbers were obtained from  a fit which  used  the unitarity triangle approach, formalism that   is supposed to take properly into account the unitarity property. Unfortunately this does not happens, the second form for $\cos\delta$ providing an imaginary value. As we will show in the next section this is a characteristic of this fit,  not an accident.
The corresponding values for $\cos\delta$  are different because, e.g., in the first case, the parameters, $a,\,b,\,c,\,d,\, e$ and $ f$ do not come from the same doubly stochastic matrix,  e.g., $f\ne\sqrt{1-d^2-e^2}$, or numerically, $0.0416\ne 0.1768$,  and so on. Thus even in the case when  the relations (\ref{sto}) are exactly satisfied, by choosing four independent moduli and constructing with them a double stochastic matrix, the numerical results show that unitarity could not be  satisfied, many  $\cos\delta$ values being not physical, and these values  depend on the {\em chosen four independent moduli}, although the solution for $\cos\delta$ of Eqs.\,(\ref{unitary}) must be unique. Hence we have to find a solution to this problem.

\subsection{Unitarity triangle method}

The second phenomenological model is defined by the orthogonality relations of rows, and, respectively columns of a unitary matrix, {\em together} with the double stochasticity relations (\ref{sto}). The last condition was never  used  in the previous approaches, see Ref.\cite{BLO}, or \cite{BaB}.
 Although there are six such relations usually one considers only the orthogonality of the first and the third columns  of $U,$   relation that is written as
\begin{eqnarray}
U_{ud} U_{ub}^* + U_{cd} U_{cb}^* + U_{td} U_{tb}^*=0\label{ort}
\end{eqnarray}
where $*$ denotes the complex conjugation.
The above relation can be visualised as a triangle in the complex plane, and usually it  is  scaled by dividing 
  through the middle term such that the length of one side is 1.  In fact one may divide by any other term because all the three possible triangles are {\em similar}, i.e. they have the {\em same angles}. On the other hand  all the six triangles obtained from all orthogonality relations such as (\ref{ort}) are {\em equivalent}, i.e. they have the same {\em area}, $A$, and furthermore the relation $J=2\,A$ holds, where $J$ is the Jarlskog invariant, see \cite{JS}-\cite{BL}. The side lengths of these triangles are auxiliary parameters, and have no physical significance. In contradistinction to  sides, the angles are measurable quantities such that they are important from an experimental point of view, see in this respect Ref. \cite{BaB}. By taking into account the relation (\ref{toy1}), 
the other sides of the triangle (\ref{ort}) have the lengths
\begin{eqnarray}
R_b^{(1)}&=&\left|\frac{U_{ud} U_{ub}^*}{U_{cd} U_{cb}^*}\right|=\frac{c\,\sqrt{1-b^2-c^2}}{d\,f},\label{tri1}\\
R_t^{(1)}&=&\left|\frac{U_{td} U_{tb}^*}{U_{cd} U_{cb}^*}\right|=\frac{\sqrt{b^2+c^2-d^2}\,\sqrt{1-c^2-f^2}}{d\,f}\nonumber
\end{eqnarray}
and, respectively,
\begin{eqnarray}
R_b^{(2)}&=&\left|\frac{U_{ud} U_{ub}^*}{U_{cd} U_{cb}^*}\right|=\frac{a\,\sqrt{1-a^2-b^2}}{d\,\sqrt{1-d^2-e^2}},\label{n2}\\
R_t^{(2)}&=&\left|\frac{U_{td} U_{tb}^*}{U_{cd} U_{cb}^*}\right|=\frac{\sqrt{1-a^2-d^2}\,\sqrt{a^2+b^2+d^2+e^2-1}}{d\,\sqrt{1-d^2-e^2}}\nonumber
\end{eqnarray}
 Since the exact fulfilment of Eqs.(\ref{sto}) does not mean the fulfilment of the unitarity property, as our numerical results (\ref{u1}-\ref{u2}) show, that property is implemented in this approach by the  conditions: all  ratios should be positive, $R_a\ge 0,\,a=b,\,\,t,\dots $, and the following inequalities should be satisfied \cite{BL}
\begin{eqnarray}
|R_b^{(i)} - R_t^{(i)}|\le 1 \le R_b^{(i)} + R_t^{(i)},\quad i=1,2,\dots\label{tri2}\end{eqnarray}
that are  equivalent with the conditions (\ref{unit1}). Indeed  by computing the above ratios  by using matrices $M_1$ and $M_2$ one gets 
\begin{eqnarray}R_b^{(1)}&=&0.438,\quad R_t^{(1)}=0.506,\label{n7}\\
R_b^{(2)}&=&1.156,\quad R_t^{(2)}=0.438\nonumber\end{eqnarray}
such that the relations (\ref{tri2}) give
\begin{eqnarray}
0.068\le 1\le 0.944,\quad 0.718\le1\le 1.594\label{n3}\end{eqnarray}
showing the perfect (theoretical) equivalence at this level of both the phenomenological models. If now we compute the above ratios by using moduli from the fit \cite{JC1} one finds
\begin{eqnarray}R_b^{(1)}&=&0.377,\quad R_t^{(1)}=0.83,\label{n5}\\
R_b^{(2)}&=&0.613 i,\quad R_t^{(2)}=0.404\nonumber\end{eqnarray}
in perfect agreement with the results (\ref{n4}), showing the inconsistency of the fit.

The Eqs.(\ref{tri1}-\ref{n2}) represent the correct form for the side lengths of  the ``standard'' unitarity triangle. Its angles are easily related to the phases of $U_{cd}$ and  $U_{td}$.

The other unitarity triangle which will provide another two independent angles is given by the orthogonality of the second and the third columns
\begin{eqnarray}
U_{us} U_{ub}^* + U_{cs} U_{cb}^* + U_{ts} U_{tb}^*=0\label{ort1}
\end{eqnarray}

One gets similarly
{\small\begin{eqnarray}
R_{c}^{(1)}&=&\left|\frac{U_{us} U_{ub}^*}{U_{cs} U_{cb}^*}\right|=\frac{b\,c}{f\,\sqrt{1-d^2-f^2}},\label{tri4}\\
R_{s}^{(1)}&=&\left|\frac{U_{ts} U_{tb}^*}{U_{cs} U_{cb}^*}\right|=\frac{\sqrt{1-c^2-f^2}\,\sqrt{d^2+f^2-b^2}}{f\,\sqrt{1-d^2-f^2}}\nonumber
\end{eqnarray}}
and, respectively
\begin{eqnarray}
R_{c}^{(2)}&=&\left|\frac{U_{us} U_{ub}^*}{U_{cs} U_{cb}^*}\right|=\frac{b\sqrt{1-a^2-b^2}}{e\,\sqrt{1-d^2-e^2}},\label{tri5}\\
R_{s}^{(2)}&=&\left|\frac{U_{ts} U_{tb}^*}{U_{cs} U_{cb}^*}\right|=\frac{\sqrt{1-b^2-e^2}\,\sqrt{a^2+b^2+d^2+e^2-1}}{e\,\sqrt{1-d^2-e^2}}\nonumber
\end{eqnarray}
Computing the above ratios with the same numbers as in (\ref{n5}) one gets

\begin{eqnarray}R_c^{(1)}&=&0.02,\qquad R_s^{(1)}=1.01,~~\label{n6}\\
R_c^{(2)}&=&0.038 i,\quad ~R_c^{(2)}=1.01\nonumber\end{eqnarray}
showing the same phenomenon as that given by the relations (\ref{n5}) concerning the  implementation of unitarity constraints for  this triangle.

Similar to the first case, two angles of the  triangle generated by the relation (\ref{ort1}) are directly connected to the phases of the elements $U_{cs}$ and $U_{ts}$.  From the above relations we may obtain all the angles of the two triangles generated by the relations (\ref{ort}), and, respectively, (\ref{ort1}).

For getting the relationship between the phases of complex entries of the CKM matrix (\ref{ckm}) and the moduli and phase  $\delta$ we make the following notation $U_{kl}=\pm |U_{kl}|e^{i\,\omega_{kl}} $, $k=2,3,\,\,l=1,2$, where the minus sign will be taken in front of $|U_{21}|$ and $|U_{32}|$. We make full use of the property   that in our parameterisation (\ref{ckm}) of the CKM matrix, the entries
$U_{ud},\,\,U_{us},$ $\,\,U_{ub},\,\, U_{cb}$ and $U_{tb}$ are all real quantities, and of the invariance of unitarity triangles angles  with respect to the scaling factor. Thus we write  the relations (\ref{tri1}) in a complex form as
\begin{eqnarray}
\frac{\cos \omega_{21} + i \sin\omega_{21}}{R_b}& =& - \,\frac{U_{cd}\,U_{cb}^*}{U_{ud}\,U_{ub}^*}\,,\label{l15}\\
\frac{(\cos \omega_{31}+ i\sin\omega_{31})\,R_t}{R_b} &=&  \, \frac{U_{td}\,U_{tb}^*}{U_{ud}\,U_{ub}^*}\nonumber\end{eqnarray}
 Please remark that the phases $\omega_{21}$ and $\omega_{31}$ coincide with the phase of $-U_{cd}$, and, respectively, $U_{td}$. The above phases coincides, in some cases modulo $\pi$,  with the angles of the unitarity triangle obtained by scaling the relation (\ref{ort}) dividing by the first term, i.e. $U_{ud}\,U_{ub}^*$. Since the triangles are similar the following relations hold
$\omega_{21}=\gamma,\,\, \omega_{31}=\alpha$.
From the first relation (\ref{l15}) we easily obtain
\begin{eqnarray}
\sin\omega_{21}&=&\frac{s_{12}\,s_{23}\,c_{23}\,\sin \delta\,R_b}{s_{13}\,c_{12}}\,,\label{l50}\\
\cos\omega_{21}&=&\frac{s_{23}(s_{13}\,s_{23}\,c_{12}+ s_{12}\,c_{23}\cos \delta)R_b}{s_{13}\,c_{12}}\nonumber\end{eqnarray}
where from one gets
\begin{eqnarray}\tan\omega_{21}=\frac{s_{12}\,c_{23}\,\sin \delta}{s_{13}\,s_{23}\,c_{12} + s_{12}\,c_{23}\,\cos \delta}\label{l17}\end{eqnarray}
The above formula for $\tan\omega_{21} $ depends only on theoretical parameters entering (\ref{ckm}), and  does not depend on the lengths of the unitarity triangle. It shows that it is not necessary the use of  unitarity triangles construction to obtain their angles, they can be easily obtained from the moduli $V_{ij}$, through relations such as (\ref{coef}) and (\ref{cod}). 

From  relations (\ref{l50}) we get a direct formula for $\cos \delta$. Indeed, from the identity
$\sin^2 \omega_{21} +\cos^2\omega_{21} =1$
we find
\begin{eqnarray}
\cos \delta = \frac{s_{13}^2\,c_{12}^2 - s_{23}^2 (s_{12}^2\, c_{23}^2+s_{13}^2\,s_{23}^2\,c_{12}^2)R_b^2}{2\,s_{12}\,s_{13}\,s_{23}^3\,c_{12}\,c_{23}}
\label{c1}\end{eqnarray}
that depends on $s_{ij}$ and $R_b$. If in it we substitute the formulae for $s_{ij}$ from relations  (\ref{coef}), and
 $R_b$ from relations  (\ref{tri1}), we find the expression given in Eq.(\ref{coef}) for $\cos \delta$, that shows at this level the perfect equivalence between the two approaches. Hence there are  explicit formulae for $\cos\delta$ in terms of triangle lengths and  mixing angles $s_{ij}$.

Doing  similar calculations by starting with the second Eq.(\ref{l15}) one gets

\begin{eqnarray}
\tan\omega_{31}&=&\frac{s_{12}\,s_{23}\,\sin\delta}{-s_{13}\,c_{12}\,c_{23}+s_{12}\,s_{23}\,\cos\delta}\label{l16}\end{eqnarray}
and, respectively,

\begin{eqnarray}
\cos \delta = \frac{-s_{13}^2\,c_{12}^2\,R_{t}^2/R_b^2 + c_{23}^2 (s_{12}^2\, s_{23}^2+s_{13}^2\,c_{23}^2\,c_{12}^2)}{2\,s_{12}\,s_{13}\,s_{23}\,c_{12}\,c^3_{23}}\label{c2}\end{eqnarray}

In conclusion the orthogonality relation (\ref{ort}) provides only  two independent angles, that depend on  {\em four} independent parameters, $s_{12},\,\,s_{13},\,\,s_{23},$ and $\delta$, and for a complete determination  of parameters entering a unitary matrix  we have to use another unitarity triangle, for example that generated by the relation (\ref{ort1}). By doing similar computations one gets
\begin{eqnarray}
\tan\omega_{22}&=&\frac{c_{12}\,c_{23}\,\sin\delta}{-s_{12}\,s_{13}\,s_{23}+c_{12}\,c_{23}\,\cos\delta}\,, \\
\tan\omega_{32}&=&\frac{c_{12}\,s_{23}\,\sin\delta}{s_{12}\,s_{13}\,c_{23}+c_{12}\,s_{23}\,\cos\delta}
 \end{eqnarray}

 Hence we can write the matrix (\ref{ckm}) under de form 
\begin{eqnarray}
U=\left(\begin{array}{ccc}
U_{ud}&U_{us}&U_{ub}\\
-|U_{cd}|e^{i \omega_{21}}&|U_{cs}|e^{i \omega_{22}}&U_{cb}\\
|U_{td}|e^{i \omega_{31}}&-|U_{ts}|e^{i\omega_{32} }&U_{tb}\end{array}\right)\label{unghi}\end{eqnarray} 
and we call $\omega_{ij}$ the fundamental phases since with them one gets  the angles of 
{\em all} unitarity triangles.  The phases entering the matrix (\ref{unghi}) corresponding to the moduli matrix $M_2$ are given by
\begin{eqnarray}
\omega_{21}&=&99.46^{\circ},\quad\omega_{31}=121.427^{\circ},\\
\omega_{22}&=&99.50^{\circ},\quad\omega_{32}=98.153^{\circ}\end{eqnarray}

Our proposal is to use the matrix form (\ref{unghi})  in all phenomenological analyses, and the experimenters have to measure all its phases which  really appear as parameters in many electroweak processes. For a similar proposal see Ref.\,\cite{AKL}.

The  important conclusions of this subsection are: a)  for a complete determination of a unitary matrix from data when one uses the unitarity triangle model one needs   the use of at least two unitarity triangles if one wants to obtain reliable results, b) the use of double stochasticity relations (\ref{sto})  is compulsory, otherwise the fit gets senseless,   and c) checking of fit results to see if them are invariant with respect to the choice of four independent moduli used to parametrise the data. The numerical results, provided by the relations (\ref{n4}), (\ref{n5}) and  (\ref{n6}), show that all the above requirements were not fulfilled by  the fit \cite{JC1}.

 \subsection{Recovery of unitary matrices from error affected data}
Until now  we supposed that the double stochasticity relations 
(\ref{sto}) were exactly satisfied, i.e. the data were not affected by errors. Even in this case by using {\em measured}, or {\em fit determined}  moduli, and  by forming with them double stochastic matrices we got that the numerical computations on $\cos\delta$ and  lengths of unitarity triangles lead to unphysical results, see Eqs.\,(\ref{n1}-\ref{n4}),  (\ref{n7}-\ref{n5}) and (\ref{n6}).  By consequence we have to see how the reconstruction algorithm which works for data coming from exact double stochastic matrices  has to be modified in order to provide reliable results in the presence of errors.  The experimental data on moduli do not satisfy the double stochasticity relations (\ref{sto}), hence the first condition which  must be imposed when doing a fit is that these relations should be satisfied with a great precision, otherwise one cannot speak of unitarity fulfilment, because    the recovery process gets senseless. Since the double stochastic matrices obtained from data by using different groups of four independent moduli are really different,  leading to different values for $\cos\delta$, see e.g. Eqs.(\ref{n1}-\ref{n4}), we have to impose that all $\cos\delta$ values should be (approximately) the same. On the other hand the explicit form for  $\cos\delta$ depends on the four independent  moduli  we choose to parameterise the data,  hence in analysing the data we have to make full use  of the axiom 2a) from Introduction.

The recovery method of unitary matrices from data that we expose in the following is a least squares method for checking the compatibility of  data with the theoretical models in both the approaches,  and stresses the necessity that the $\chi^2$-function have to contain two kind of terms: the first has to impose the fulfilment of unitarity constraints, and the second should take into account the physical quantities measured in experiments. The piece  of unitarity constraints has an independent part provided by the double stochasticity relations, and a  dependent one  upon the phenomenological model. Constraints implied by  double stochasticity relations are written as 
\begin{eqnarray}
&&\chi^2_{ds}=\label{chi1}\\
&&\sum_{j=u,c,t}\left(
\sum_{i=d,s,b}V_{ji}^2-1\right)^2
+\sum_{j=d,s,b}\left(
\sum_{i=u,c,t}V_{ij}^2-1\right)^2\nonumber
\end{eqnarray}
The constraints generated by the unitarity condition method are given by
\begin{eqnarray}
\chi^2_1=\sum_{i < j}(\cos\delta^{(i)} -\cos\delta^{(j)})^2,\,\,\,\,-1\le\cos\delta^{(i)}\le 1\label{chi1a} 
\end{eqnarray}
A similar formula one gets  for the second phenomenological model where instead of $\cos\delta$ one uses the side lengths $R_a$, see Ref.\cite{Di}.
 The third component 
which takes into account the experimental data has the form
\begin{eqnarray}
\chi^2_2=\sum_{i}\left(\frac{d_{i}-\widetilde{d}_{i}}{\sigma_{i}}\right)^2\label{chi2} 
\end{eqnarray}
where  $d_{i}$ are the theoretical functions one wants to be found from fit, $\widetilde{d}_{i}$ is  the numerical matrix that describes the corresponding experimental data, while  $\sigma$ is the matrix of errors associated to $\widetilde{d}_{i}$. 
 Hence a $\chi^2$-test could be the function
\begin{eqnarray}
\chi^2=\chi^2_{ds}+\chi^2_1 +\chi^2_2\label{fit}\end{eqnarray}
 which  will be used in numerical computations. 

 The compatibility condition of Eqs.\,(\ref{unitary}), together with the invariance of physical quantities with respect of the four independent moduli group chosen to parametrise the data,  imply strong conditions on all the moduli given by the relations $\cos\delta^{(i)} \approx\cos\delta^{(j)}$ for $i\ne j$, see Eq.\,(\ref{chi1a}), which are not easy to implement in a fit, but gives us at least one reward: a method for doing statistics on unitary matrices.

It is well known that the problem of doing statistics on (moduli) of unitary matrices was an open problem, see e.g. \cite{PS}.
The embedding relations (\ref{ds1}) suggest that the right quantities to look at them are  the square moduli, $|U_{ij}|^2$. Indeed the convexity property of  double stochastic matrices implies that 
if we have a set of  unitary matrices $U_1,\,\dots,U_n$ then
\begin{eqnarray}
M^2&=&\sum_{i=1}^{i=n} \,x_i \cdot |U_k|^2,\label{db},\qquad \sum_{i=1}^{i=n} \,x_i=1,\\
&& 0\le x_i \le 1,\,\quad i=i,\dots,n \nonumber
\end{eqnarray}
is a double stochastic matrix. In other words we have to do statistics on the set of double stochastic matrices generated by unitary ones. Thus this property allows us to calculate correctly mean values, $\langle M\rangle$, and error matrices, $\sigma_M$, for a set of double stochastic ones, and by consequence for unitary matrices, as follows
\begin{eqnarray}
\langle M\rangle&=&\sqrt{\left(\sum_{k=1}^{k=n} \, |U_k|^2\right)/n}\,\,,\label{db1}\\
\sigma_M&=&\sqrt{\left(\sum_{k=1}^{k=n} \, |U_k|^4\right)/n\,\, -\,\,\langle M \rangle^4}\nonumber
\end{eqnarray}
 where the square roots are taken entry wise. If the mean value matrix obtained in this way is not too far  from a unitary one, one can reconstruct from it an (approximate) unitary matrix. Also important is the relationship between the ``central'' values matrix and those obtained by adding the error matrix $\sigma_M$.
Thus the above formulae suggest that the $\pm 3\,\sigma$ matrices should be calculated by the formulae
\begin{eqnarray}
M_+=\sqrt{M^2+3  \sigma_M^2},\quad  M_-=\sqrt{M^2-3  \sigma_M^2}\end{eqnarray} 
In this way we provided a complete  formalism for doing fits on data from the electroweak sector, which is a very robust one such as can be   seen in the next Section.
\section{Testing data and fits}

In the following we want to show what are the main consequences of the above formalism, especially what are the subtleties and the results obtained when doing statistics on unitary matrices by the above method.  For  that we will make use of all 165 different forms for $\cos\delta$. We test the Reviews of Particle Physics 2004 \cite{PDG} and 2006 data \cite{WM}, the lattice computations of CKM moduli \cite{Ok}, and the published fits \cite{JC} and \cite{JC1} . 

1. 2004  PDG data \cite{PDG}. As it is well known there one finds  upper and lower bounds for each modulus, at 95\% confidence level.  We used the mean values, calculated as half the sum of upper and lower bounds, and the corresponding $\sigma_i$ computed as half the difference of the bounds. One gets that the double stochasticity is quite well satisfied by the mean values, all the six  sums differing from 1 by amounts of order $10^{-6} - 10^{-4}$, which lead to
\begin{eqnarray}
\langle\cos\delta\rangle= 0.063+0.008 i
\end{eqnarray}
Because the mean real and imaginary parts are small, the above results suggest that the central data are consistent with a $\delta$ value around $90^{\circ}$, and a fit done by  using our method provided the numerical matrix
\begin{eqnarray}
V_1=\left(\begin{array}{ccc}
0.9748379&0.222885&0.00365667\\
0.222699&0.974025&0.0409987\\
0.0098077&0.0399759&0.999153\end{array}\right)
\end{eqnarray}
which leads to the following value
\begin{eqnarray}\delta=(90.01\pm 0.41)^{\circ} \end{eqnarray}
For the above matrix the double stochasticity property is well satisfied, all  six relations (\ref{sto}) taking values whose  magnitude is of the  order of $10^{-6}$, and this implies a small (statistical) error for $\delta$.

If one uses the recommended values \cite{PDG} for the entries on the first two rows, and the mean values for moduli on the third row one gets big non-physical values
\begin{eqnarray}
\langle\cos\delta\rangle=-24.15-8.16 i,\,\,
 \sigma_{\cos\delta}=504.21-0.39 i
\end{eqnarray}
which show that the recommended values are quite far from values compatible with unitarity requirements.

2. 2006  PDG data \cite{WM}. We made a data   modification for one single modulus, $V_{tb}$, by taking its central value equal to  $V_{tb}=0.99912$, instead of $V_{tb}=0.77^{+0.18}_{-0.24}$, since we saw that the obtained results are  similar, and one gets
\begin{eqnarray}
\langle\cos\delta
\rangle=21.557-21.320 i, \,\,
\sigma_{\cos\delta}=62.738+7.326 i~~
\end{eqnarray}
which shows that the mean data are far from being  compatible with unitarity.
A fit with our formula (\ref{fit}) around the central values from PDG 2006 data provides the matrix
\begin{eqnarray}
V_2=\left(\begin{array}{ccc}
0.974290&0.225260&0.004114\\
0.225064&0.973579&0.038602\\
0.010256&0.037441&0.999246\end{array}\right)\label{v2}
\end{eqnarray}
 where from we get
\begin{eqnarray} \delta=(101.18\pm 0.18)^{\circ}\end{eqnarray}

The 2006  PDG data \cite{WM} bring an innovation: entering of global fits results into Review of Particle Physics book, see Ceccucci et al. contribution to CKM quark-mixing matrix \cite{CLS} in the book. The fits taken into account are \cite{JC} and \cite{Bo1}.
By using the central values provided there one finds
\begin{eqnarray}
\langle\cos\delta\rangle=0.5084, \quad \sigma_{\cos\delta}=0.1002
\end{eqnarray}
which is equivalent to 
\begin{eqnarray} \delta=(59.44^{+6.46}_{-6.93})^{\circ}\end{eqnarray}
which shows at the level of $1 \sigma$ an important statistical  spreading caused by a poor fulfilment of the double stochasticity relations for the first row and the first column. If we look at the central values $V_c+\sigma$, and respectively $V_c-\sigma$, where $+\sigma $ and  $-\sigma $ are slightly different, see Eq.(11.26) in \cite{CLS}, we get
\begin{eqnarray}
\langle\cos\delta^+\rangle&=&1.4175+0.0508 i, \\
 \sigma_{\cos\delta^+}&=&6.1236 - 0.0118 i
\end{eqnarray}
and respectively
\begin{eqnarray}
\langle\cos\delta^-\rangle&=&0.3076+0.2324 i,\\
 \sigma_{\cos\delta^-}&=&6.2063 - 0.0115 i
\end{eqnarray}
Both results show that moduli values  $V_c\pm\sigma $ have no physical relevance and provide one numerical example showing the shortcomings of the unitarity triangle approach such as it is used in the present day. The big values for $\sigma_{\cos\delta^{\pm}} $ confirm that many $\cos\delta$ values take values outside the physical range $(-1,\,1)$, and an important number takes imaginary values.
For example in the second case,  $V_c-\sigma$, the real values are within the interval $\cos\delta \in(-16.45,\, 19.17)$, and the imaginary ones, 19 values from 165, within the interval
$\cos\delta \in(-0.355,\, 7.525)i$

 In our opinion this example shows clearly the  two causes which generate such results: a) fitting only with a single unitarity triangle and  using essentially only three moduli, $b,\,c$ and $f$,  and b) non existence of a sound procedure for doing statistics on moduli of unitary matrices.

3. By using the lattice results from Ref.\,\cite{Ok} on CKM matrix moduli one gets
\begin{eqnarray}
\langle\cos\delta\rangle=8.962-11.315 i,\,\,
\sigma_{\cos\delta}=20.15+5.03 i \end{eqnarray}
which shows that lattice computations are still  far from  results compatible with unitarity constraints. The above results show that the double stochasticity property is not satisfied, hence the numerical effort has to be done to improve it.
By looking for a compatible unitary matrix around the mean  moduli values  from Ref.\,\cite{Ok}, with our method one gets the numerical matrix
\begin{eqnarray}
V3=
\left(\begin{array}{ccc}
0.974330&0.225086&0.0041314\\
0.225003&0.973562&0.0393922\\
0.007401&0.038911&0.999215\end{array}\right)\label{latt}
\end{eqnarray}
which leads to $\delta=(56.04\pm 0.62)^{\circ}$ . This moduli matrix  can be used as a good ``unitary witness'' to improve the numerical algorithms used to obtain  results as those given in  Ref.\,\cite{Ok}. 

The sensitivity of our method to small moduli variations can be seen by looking at the entries of matrices (\ref{v2}) and (\ref{latt}). The difference of the  moduli on the first two rows is of the order of $10^{-4}$, while  $\delta$ changes from $101^{\circ}$ to $56^{\circ}. $
\vskip2mm
4. CKM fitter Group 2001 results \cite{JC1}.
The real and imaginary  values for $\cos\delta$ are within the bounds 
$\cos\delta\in (-1.082, 1.596)$, respectively, $ \cos\delta\in i(-2.431, 0.835)$ and one gets
\begin{eqnarray}
\langle\cos\delta\rangle=0.373-0.087\, i,\,\,
\sigma_{\cos\delta}=0.358+0.907\, i \end{eqnarray}
 Hence one could say that the central values together with their $\pm\sigma$ companions are not compatible with unitarity. This happened because the invariance of the physical results with respect to all the choices for the {\em four} independent moduli groups, and the double stochasticity relations  were not yet implemented in the unitarity triangle model,  and, by consequence, the fit have no physical significance.

5. CKM fitter Group 2005 results \cite{JC}. The last published results by  CKM fitter Group  are, in our opinion,  the best published results until now. Thus our fourth selected  matrix is given by the central values from \cite{JC}, see their Table 3
\begin{eqnarray}
V_4=\left(\begin{array}{ccc}
0.97400&0.2265&0.00387\\
0.2264&0.97317&0.04113\\
0.00826&0.04047&0.999146\end{array}\right)\label{di2}
\end{eqnarray}
The errors on moduli are of the order $ 5.5\times 10^{-4}$ for $V_{ud}$ and  $V_{cs}$,
and of order $ 2.5\times 10^{-3}$ for $V_{us}$ and  $V_{cd}$. The double stochasticity is well satisfied obtaining numbers of the order $( 10^{-8} -  10^{-6})$. On this matrix one can check how the fulfilment degree of double stochasticity property reflects in $\cos\delta$ values. One gets  for $\delta$ and  Jarlskog invariant, $J$, the following values 
\begin{eqnarray}\delta=(62.07^{+ 3.79}_{-3.92})^{\circ},\quad J=(3.84^{+0.62}_{-0.79})\times 10^{-5}\end{eqnarray}
We remind  that for a double stochastic matrix the statistical error  of $\delta$ is zero, such that the errors for $\delta$ give a measure of the double stochasticity fulfilment. As a curiosity we remark that, from the point of view of unitarity constraints,  the numerical results (\ref{di2}) are better than those appearing in PDG 2006 data, \cite{CLS},  perhaps since the last ones were obtained by  merging results from two different fitting groups. Unfortunately the errors accompanying the matrix (\ref{di2}) have no physical relevance if are used according with the present day rules.

By doing a fit with our method around the values from Eq.(\ref{di2}) one gets such a matrix, which is only slightly different from the above one.
If we check now the results for $V_4+1\sigma$, and $V_4-1\sigma$, where the matrices  $+\sigma$  and  $-\sigma$ are different, see their Table 3, and one adds the errors such as they are given,  one gets
\begin{eqnarray}
\langle\cos\delta\rangle_+&=&-2.214-4.559 i,\\
\langle\cos\delta\rangle_-&=&-2.91-5.33i\nonumber\end{eqnarray}
If we proceed by using our method, i.e. we calculate $\delta_{\pm}$  by using the moduli obtained with the formulae  $\sqrt{V_4^2\pm\sigma^2}$ one gets
\begin{eqnarray}
\delta_+=(61.02^{+7.16}_{-7.7})^{\circ} \quad {\rm and}  \quad\delta_-=(61.91^{+3.45}_{-3.56})^{\circ}\end{eqnarray}
Also one finds that the $+\sigma$ error matrix   is overestimated since, for example,   computations with the moduli matrix 
 $\sqrt{V_4^2+3\sigma^2}$ lead to complex values for  $\langle\cos\delta\rangle$, namely 
$\langle\cos\delta\rangle=0.554 - 0.015 i$.

 6. Other selected matrices.   Around the central values from PDG 2006 data we imposed a supplementary constraint, namely,   $V_{td}/V_{ts}=0.208 $, and we got
\begin{eqnarray}
V_5=\left(\begin{array}{ccc}
0.973556& 0.228407& 0.00431\\
0.228289& 0.972704& 0.0416 \\
0.00851678& 0.0409463& 0.999125 \end{array}\right)
\end{eqnarray}
and  from it one finds $ \delta=(63.706\pm 0.228)^{\circ}.$

More exotic matrices are: a) around central PDG 2004 data values
\begin{eqnarray}
V_6=\left(\begin{array}{ccc}
 0.975282& 0.220925& 0.00340635 \\
 0.220702& 0.974397& 0.04299 \\
 0.0106931& 0.0418558& 0.999066 \\
  \end{array}\right)
\end{eqnarray}
with
\[ \delta=(98.1\pm 6.4)^{\circ}\]
b) around PDG 2006 data central values

\begin{eqnarray}
V_7=\left(\begin{array}{ccc}
 0.97328& 0.2295& 0.0062 \\
 0.229112& 0.97256& 0.04044 \\
 0.0153& 0.037939& 0.99916 \\
  \end{array}\right)
\end{eqnarray}
which leads to 
\[ \delta=(155.7\pm 13.5)^{\circ}\]
or c)
\begin{eqnarray}
V_8=\left(\begin{array}{ccc}
 0.973514& 0.228585& 0.004311 \\
 0.228564& 0.97264& 0.0416 \\
 0.005316& 0.041484& 0.999125 \\
  \end{array}\right)
\end{eqnarray}
equivalent  to 
\[ \delta=(6.4\pm 4.0)^{\circ}\]
The double stochasticity property is not very well satisfied by the last three exotic matrices. We provided them  to see how our method for doing statistics on unitary matrices works. However,
from a phenomenological  point of view, all the above eight matrices are good, and each one leads to a perfectly acceptable unitary matrix. The above examples show that the obtained unitary matrices depend strongly on the numerical values around one looks for a compatible unitary matrix. Changing them a little bit one gets different values for  $\delta$. Hence the data are not known with a sufficient  precision in order to tighten significantly the $CP$ violating phase $\delta$. 

The novelty brought by the embedding of unitary matrices into the convex set of double stochastic matrices is the following: we can use
 the double stochasticity property, and from the (approximate) double stochastic matrices $V_i,\,\, i=1\dots,9$, where $V_9$ denotes  the approximate numerical form of the  $M_2$ matrix, we get a continuum of  (approximate) double stochastic matrices
\begin{eqnarray}
W^2&=&\sum_{i=1}^{i=8}x_i \, V_i^2+(1-\sum_{i=1}^{i=8}x_i)V_9\,,\label{cont}\\
&& 0\le x_i\le 1,\,\,i=1,\dots,8,\quad \sum_{i=1}^{i=8}x_i\le 1\nonumber\end{eqnarray}
continuum which depends on eight arbitrary parameters $x_i$. Its consequence is that with  the $W$ matrix one can obtain practically any value for $\delta$ within the interval $(0^{\circ}, 180^{\circ}) $, all of them being relatively  on the same footing in what concerns their agreement with the experimental data on the moduli $V_{ij}$, and with the fulfilment of unitarity constraints. 

If we use the formulae (\ref{db1}) to do statistics on the above nine matrices
 we find at the symmetric point $x_i=1/9,\,i=1,\cdots,8,$  the matrix
\begin{eqnarray}
W&=\left(\begin{array}{ccc}
0.974134& 0.225921,&0.0042982 \\
 0.225766& 0.973323& 0.0409179 \\
0.009583& 0.040019& 0.999153 \end{array}\right)
\end{eqnarray}
 and its associated  error matrix 
\begin{eqnarray}
\sigma=\left(\begin{array}{ccc}
 0.00119& 0.00118& 7.4\times 10^{-6}\\
0.00118& 0.00117& 1.0\times 10^{-4} \\
5.7\times 10^{-5}&11.68\times 10^{-5} &9.88\times 10^{-5} \end{array}\right)\label{a100}\end{eqnarray}
which provide  the following value for $\delta$
\begin{eqnarray}\delta=(81.12 \pm 1.23)^{\circ}\label{d1}\end{eqnarray}
Although the moduli matrices, $V_i$, lead to $\delta$ values practically on all the physical interval, $(0,\,180)^{\circ}$, the mean value $W$ satisfies the double stochasticity property quite well shown by the relatively small statistical error in formula (\ref{a100}), which leads to a small error for $\delta$, see Eq.(\ref{d1}).
Of course the true matrix $\sigma$ has to be calculated by taking into account all  $\sigma_i$ matrices, by using the same property: one takes the mean of squared $\sigma_i^2$, which lead to  a new matrix whose entries are a  little bit bigger  than those appearing in (\ref{a100}).

The above results show that the present day fits are not able to determine the CKM matrix entries with a high precision, and by consequence neither $\delta$,  nor the $J$ invariant can be   well defined.     The  computations, as well as all the numbers obtained by the so called global fits, use only  partial information coming from experiments, which has as consequence the existence of a continuum of unitary matrices compatible with a set of experimental data  and unitarity constraints.  As the numerical computations show this type of global fit gives better results and in the same time it provides a  sound method for doing statistics on unitary matrices.

\section{Conclusion}

In the paper we have shown that the compatibility between the moduli data 
and the theoretical model (\ref{ckm}) can be obtained if and only if three
 conditions are simultaneously satisfied: a) the double stochasticity 
relations (\ref{sto}) hold to a level of at least $10^{-6}$, 
 b) the physical condition $-1\le \cos\delta \le 1$ is satisfied,
 and c) the numerical values for $\cos\delta$ do not depend on the four
 independent moduli chosen to parametrise the data.

Our numerical checking of the two approaches in Section 2 has shown that
 both the theoretical models, unitarity triangle method and the unitarity 
condition method, send the same signal concerning the compatibility of
  data with  unitarity requirements.
 Thus now we have two phenomenological approaches, and both of them have to
 give results consistent each other.

 The numerical implementation of the $\chi^2$-test, such as (\ref{fit}),
 is a little bit more complicated  in both the approaches  when one  makes
 use of the exact form of unitarity constraints. This happens since by 
looking at  the denominators of the relations (\ref{coef}-\ref{cod}),
 (\ref{tri1}-\ref{u2}), (\ref{tri4}-\ref{tri5}), and those similar to them,
 one sees that there appear square roots which by modifying  a little bit
 the moduli values, change easily their values from real to imaginary ones.
 On the other hand by using approximations, as those used in the standard 
form of unitarity triangle model, one easily finds non physical results.
 However, as our numerical computations from Section 3 show, 
the implementation of all the exact unitarity constraints can be done.

Our computations have shown that the phenomenologists are not yet aware
  of the constraining
 power of unitarity, this one requiring a moduli matching  to an order
 of $10^{-4}-10^{-6}$, much lower than the experimental errors, tuning
 that can not be obtained by using approximate formulae. The influence 
of errors on the final results is that in case  they are not small enough,
 the double stochasticity property provides us continuum sets of
 approximate unitary matrices. Combining these results   with a correct
 method for doing statistics on unitary matrices we have obtained a powerful 
tool for checking unitarity properties of  data that will become 
 available at the LHC machine in the next years.

\end{document}